\documentclass[pss,fleqn]{w-art}
\usepackage{times,bm,psfrag}
\usepackage{w-thm}
\theoremstyle{plain}

\theoremstyle{definition}

\usepackage[]{graphicx}
\chardef\bslash=`\\ 

\begin{document}
\DOIsuffix{theDOIsuffix}
\Volume{XX}
\Issue{}
\Month{}
\Year{}
\pagespan{3}{}
\Reviseddate{}
\Accepteddate{}
\Dateposted{}
\subjclass[pacs]{}

\title[Shannon capacity in a Raleigh model]{Fluctuations of the Shannon capacity in a Raleigh model
of wireless communication}
\author[J. St\"aring]{J. St\"aring\inst{1}}
   \address[\inst{1}]{Theoretical Physics, Physics and Engineering Physics,
   G\"oteborg University/Chalmers, Sweden}
\author[A. Eriksson]{A. Eriksson\inst{2}}
\address[\inst{2}]{Physical Resource Theory, Physics and
Engineering Physics, G\"oteborg University/Chalmers, Sweden}
\author[B. Mehlig]{B. Mehlig\inst{1}}
\date{\today}
\begin{abstract}
Using the fact that the Shannon capacity $C$ of a Raleigh model
of wireless channels is a linear statistic of the channel matrix,
we calculate its variance $\mbox{var}[C]$.  We find that the expected
value $\langle C\rangle$ of the Shannon capacity is typical
in the model considered, that is the coefficient of variation
$\sqrt{\mbox{var}[C]}/\langle C\rangle$ is small.
\end{abstract}
\maketitle
\begin{SCfigure}[2][b]
\psfrag{t3}{\hspace*{-4mm}\small$n_{\rm R}$ receivers}
\psfrag{t1}{\raisebox{-2mm}{\small$n_{\rm T}$ transmitters}}
\psfrag{t2}{\raisebox{-4mm}{\small scattering medium}}
\includegraphics[width=5cm,clip]{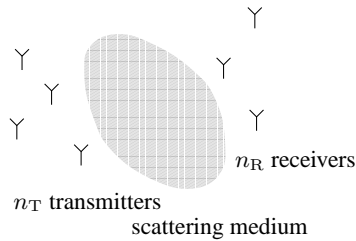}
\caption{\label{fig:0} Wireless array of $n_{\rm T}$ transmitting
and $n_{\rm R}$ receiving antennae. It is assumed that $n_{\rm T}\geq
n_{\rm R}$. The scattering medium is characterised by a 
$n_{\rm T}\times n_{\rm R}$ channel  matrix $\bm H$ with complex
matrix elements.}
\end{SCfigure}

The efficiency of a wireless channel is determined by
its Shannon capacity, $C=\log_2(1+\rho\,|H|^2 )$ where $\rho$ is
the signal-to-noise ratio and $H$ is the transfer characteristic 
of the channel \cite{fo98}. The Shannon capacity describes
the rate of information transfer (in bits per second, bps).
A crucial question in the design of multiple-anntenna 
arrays is: 
how does the channel capacity increase with the number of channels?
Consider  an array of $n_{\rm T}$ transmitters and $n_{\rm R}$ receivers
as shown schematically in Fig.~\ref{fig:0}. The scattering medium
is characterised by a $n_{\rm T}\times n_{\rm R}$ channel  matrix $\bm
H$ with complex matrix elements $H_{kl}$ determining the amplitude
of the $l$th receiving antenna arriving from transmitter $k$.
In realistic situations, the scattering medium changes as a function
of time, and so does the capacity. Typically it
fluctuates randomly; it was therefore
suggested \cite{tel95,fo98} to calculate
an average capacity as an ensemble average over random
matrices $\bm H$. In \cite{tel95,fo98} an idealised
model (called Raleigh model in the following)
was considered: $H_{kl}$ were taken to be
uncorrelated random variables with zero mean and unit variance.
This corresponds to a regular array of antennae, spaced $\lambda/2$
apart ($\lambda$ is the wave length). In this case,
the capacity is given by
\begin{equation}
\label{eq:cap}
C({\bm H}) = \log_2\det\big({\bm 1}+\frac{\rho}{n_{\rm T}}{\bm H}{\bm
H}^\dagger\big)
\end{equation}
\hspace*{-4.15mm}and its average was calculated in \cite{tel95}.
Remarkably, the average capacity was found
to scale linearly with the number of transmitters (receivers) 
for a large number of antennae.
This observation has attracted considerable attention,
and the average Shannon capacity
in related, but more general models (incorporating correlations between
the matrix elements $H_{kl}$) has been studied in great detail
\cite{shiu00,mou00,loy01,loy02}.
It was found that correlations between the matrix elements
$H_{kl}$ reduce the Shannon capacity somewhat,
but by increasing the number of antennae,
the Shannon capacity can be increased significantly.
Empirical studies have indeed shown substantial efficiency gains
for such antenna arrays \cite{ling}.

An important question is however: how typical is the expected value of
$C$? In other words, how large are its fluctuations? It has been
argued that the distribution of the capacity is a sharply
peaked function \cite{mou00}. Recently its distribution
was calculated, for the model described
in \cite{mou00}, in the limit of large $n_{\rm T}$ and $n_{\rm R}$
using the replica technique \cite{mou03}. In this limit
the distribution was found to be Gaussian, with a finite variance.

Here we calculate the fluctuations of the Shannon capacity for
the Raleigh model (\ref{eq:cap}) exactly for arbitrary
values of $n_{\rm T}$ and $n_{\rm R}$.
We use the fact that
the capacity (\ref{eq:cap}) is a {\em linear statistic}
\cite{dys}
of the channel matrix ${\bm H}$, i.e.,  it can be expressed in the form
$C = \sum_i f(\lambda_i)$ where $\lambda_i$ are the eigenvalues
of the channel matrix. The fluctuations of a linear statistic
of a random matrix are determined by the spectral $m$-point functions.
The variance, for example, is given by
\begin{eqnarray}
\label{eq:var}
\mbox{var}[C]&=& \int\!\!{\rm d}\lambda \int\!\!{\rm d}\mu\,
K_2(\lambda,\mu) 
\log_2(1+\rho\lambda/n_{\rm T})\log_2(1+\rho\mu/n_{\rm T})
\end{eqnarray}
where $K_2(\lambda,\mu) = -\langle \sum_{ij}
\delta(\lambda-\lambda_i)
\delta(\mu-\lambda_j)\rangle+d(\lambda)d(\mu)$ 
is the two-point correlation function and
$d(\lambda) = \langle \sum_i \delta(\lambda-\lambda_i)\rangle$  is the
density of states (the one-point function).
Higher moments of $C$ can be expressed in terms of higher spectral
correlation functions.

The random matrix ensemble discussed above is in fact
the so-called Laguerre ensemble \cite{bronk}.
In this ensemble, the 
$m$-point correlation function are known exactly \cite{laguerre}.
The two-point correlation function $K_2$ is usually expressed in terms of 
the two-level cluster function $T_2$,
$K_2(\lambda,\mu) = T_2(\lambda,\mu) -d(\lambda)\delta(\lambda-\mu)$,
and
\begin{eqnarray}
T_2(\lambda,\mu) &=& (\lambda\mu)^{a/2} {\rm e}^{-(\lambda+\mu)/2}
\sum_{j=0}^{n_{\rm R}-1} \frac{j!}{\Gamma(j+a+1)} L_j^a(\lambda)
L_j^a(\mu)
\end{eqnarray}
\begin{SCfigure}[2][b]
\raisebox{-7mm}{\includegraphics[width=5cm,clip]{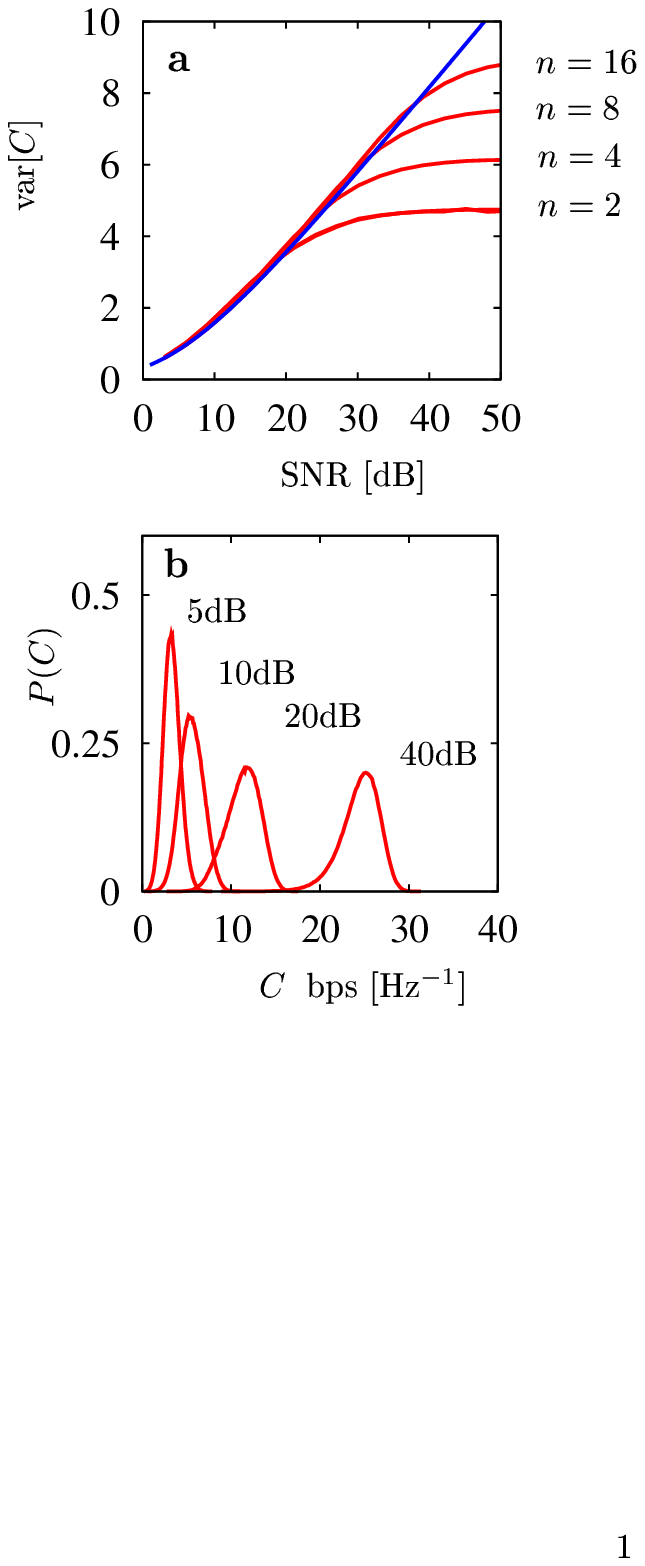}}
\caption{\label{fig:1} \textbf{a}  Variance of the channel capacity
as a function of the signal-to-noise ratio (SNR)
for $n_{\rm R}=n_{\rm T}$, with $n=2,4,8$ and $16$ (red, from bottom to top). 
Also shown is the
asymptotic result (\ref{eq:result}) 
valid in the limit $n\rightarrow\infty$ (blue).
{\bf b} Distribution of the channel capacity for $n_{\rm R}=n_{\rm
T}=n=2$ for different signal-to-noise ratios, obtained from
diagonalisations of $2\times 2$ matrices.}
\end{SCfigure}
\hspace*{-4.15mm}where $a=n_{\rm T}-n_{\rm R}$ and $L_j^a(\lambda)$ are
Laguerre polynomials \cite{grad}. 
We have evaluated the variance of the capacity using eqs. (2,3).
The results are displayed in Fig.~\ref{fig:1}{\bf a}, for the case
$n_{\rm T} = n_{\rm R} \equiv n$. Shown is the variance
of the capacity, $\mbox{var}[C]$, as a function of the
signal-to-noise ratio in the range from zero
to $50$ dB. 
For small values of $n$, the variance is found to depend
significantly on $n$, but for large values
of $n$, it appears to become independent of the number $n$ of antennae.
In the limit of large $n$, we make use of 
an asymptotic result derived in
\cite{bee} to obtain, from eqs.~(2,3)
\begin{eqnarray}
\mbox{var}[C]&\approx&\frac{1}{2\pi^2} {\cal P}
\int_0^{4n}\!\!{\rm d}\lambda
\int_0^{4n}\!\!{\rm d}\mu
\left[\frac{\mu(4 n-\mu)}{\lambda(4 n-\lambda)}\right]^{1/2} 
 \frac{c(\lambda)}{\lambda-\mu} \frac{{\rm
d}}{{\rm d}\mu} c(\mu)
\end{eqnarray}
where $c(\lambda) = \log_2(1+\rho\lambda/n)$
and ${\cal P}$ denotes the principal value.
We arrive at
\begin{eqnarray}
\label{eq:result}
\mbox{var}[C]&\approx&
\frac{1}{\pi\,(\log2)^2}\int_0^{\pi/2}\!\!\!\!\! {\rm d}\theta\,\,\, \frac{\log (1 \!+\! 4\,\rho \,{\sin^2\theta})\,\left( 1\! -\! {\sqrt{1 + 4\,\rho\, }} + 4\,\rho \,{\sin^2 \theta }
\right) }{ 1\! +\! 4\,\rho \,{\sin^2 \theta } }\,.
\end{eqnarray}
This result corresponds to a special case of the asymptotic expression
eq.~(59) in \cite{mou03} and is also shown in Fig.~\ref{fig:1}{\bf a}. 
We conclude: for small and moderate
signal-to-noise ratios, the asymptotic limit given by eq.~(\ref{eq:result})
is rapidly attained for growing values of $n$. For large
values of $\rho$, however, the convergence is seen to be much slower.
In the limit  of $n\rightarrow\infty$,  eq.~(5) together with
the result of \cite{tel95} gives
$\sqrt{\mbox{var}[C]}/\langle C\rangle\propto n^{-1}$.
But even for small values of $n$ we find 
that the coefficient of variation $\sqrt{\mbox{var}[C]}/\langle
C\rangle$ is small.
This implies that the expected value of $C$ is typical
even for a small number of antennae; the larger the signal-to-noise ratio
is, the more typical is $\langle C\rangle$.
Furthermore, since $C$ is a linear statistic, Politzer's argument
\cite{pol89} implies that the distribution of $C$  must be Gaussian
in the limit of large $n$, as noted in \cite{mou03}.
Fig.~\ref{fig:1}{\bf b}
shows that even for rather low values of $n$, the distribution
is well approximated by a Gaussian for small signal-to-noise
ratios $\rho$. For larger values of $\rho$, the distribution
is seen to develop a tail to the left. This tail, however, is found to
rapidly disappear in the limit of large $n$.


\begin{thebibliography}{22}
\bibitem{fo98} G. J. Foschini and M. J. Gans,
Wireless Personal Communications {\bf 6}, 311 (1998).
\bibitem{tel95} I. E. Telatar, Bell Lab. Report 1995.
\bibitem{shiu00} Da-Shan Shiu, G. J. Foschini, M. J. Gans, and J. M.
Kahn, IEEE Transactions on Communications {\bf 48}, 502 (2000)
\bibitem{mou00} A. L. Moustakas, H. U. Baranger, L. Balents, A. M.
Sengupta, and S. H. Simon, Science {\bf 287}, 287 (2000)
\bibitem{loy01} S. Loyka, IEEE Communication Letters, {\bf 5}, 369
(2001)
\bibitem{loy02} S. Loyka and G. Tsoulos, IEEE Communications Letters
{\bf 6}, 19 (2002)
\bibitem{ling} J. Ling, D. Chizhik, P. Wolniansky, R. Valenzuela, N.
Costa, and K. Huber, IEE Electron. Lett. {\bf 37}, 1041 (2001)
\bibitem{mou03} A. L. Moustakas. S. H. Simon, and A. M. Sengupta,
IEEE Transactions on Information Theory {\bf 49}, 2545 (2003)
\bibitem{dys} F. J. Dyson and M. L. Mehta, J. Math. Phys. {\bf 4}, 
701 (1963).
\bibitem{bronk} B. V. Bronk, J. Math. Phys. {\bf 6}, 228 (1965)
\bibitem{laguerre} T. Nagao and K. Slevin, J. Math. Phys. {\bf 34}, 2075
(1993)
\bibitem{grad} I. S. Gradshteyn and I. M. Ryzhik, Table of integrals,
series, and products, Academic Press, San Diego (1980)
\bibitem{bee} C. W. J. Beenakker,  Nucl. Phys. B {\bf 422}, 515  (1994)
\bibitem{pol89} H. D. Politzer, Phys. Rev. B {\bf 40}, 11917 (1989)
\end{thebibliography}
\end{document}